\documentclass[aps,twocolumn,superscriptaddress,altaffilletter,lengthcheck,
tightenlines,showpacs,showkeys]{revtex4-1}

\usepackage{color}
\newcommand{\be}{\begin{equation}}
\newcommand{\bn}{\begin{equation}\label}
\newcommand{\ee}{\end{equation}}
\newcommand{\bea}{\begin{eqnarray}}
\newcommand{\eea}{\end{eqnarray}}

\newcommand{\ben}{\begin{eqnarray}}
\newcommand{\een}{\end{eqnarray}}
\newcommand{\n}{\label}
\newcommand{\no}{\noindent}
\newcommand{\ga}{\gamma}
\newcommand{\Ga}{\Gamma}

\newcommand{\la}{\lambda}

\newcommand{\ro}{\rho}
\newcommand{\al}{\alpha}
\newcommand{\vphi}{\varphi}

\begin{document}
\title{Enhanced Inflation in the Dirac--Born--Infeld framework}

\author{Luis P. Chimento}\email{chimento@df.uba.ar}
\affiliation{Departamento de F\'{\i}sica, Facultad de Ciencias Exactas y Naturales,  Universidad de Buenos Aires, Ciudad Universitaria, Pabell\'on I, 1428 Buenos Aires, Argentina}
\author{Ruth Lazkoz}\email{ruth.lazkoz@ehu.es}
\affiliation{Fisika Teorikoa, Zientzia eta Teknologia Fakultatea, Euskal Herriko Unibertsitatea, 644 Posta Kutxatila, 48080 Bilbao, Spain}
\author{Mart\'{\i}n G. Richarte}\email{martin@df.uba.ar}
\affiliation{Departamento de F\'{\i}sica, Facultad de Ciencias Exactas y Naturales,  Universidad de Buenos Aires, Ciudad Universitaria, Pabell\'on I, 1428 Buenos Aires, Argentina}
\date{\today}
\bibliographystyle{plain}

\begin{abstract}
We consider the Einstein equations within  the DBI scenario for a spatially flat Friedmann-Robertson-Walker (FRW) spacetime without  a cosmological constant. We derive the inflationary scenario by applying the symmetry transformations which preserve the form of the Friedmann and conservation equations. These form invariance transformations generate a symmetry group parametrized by the Lorentz factor $\ga$. We explicitly obtain an inflationary scenario by the cooperative effect of adding energy density into the Friedmann equation. For the case of a constant Lorentz factor, and under the slow roll assumption,  we find the transformation rules for the scalar and tensor power spectra of perturbations as well as their ratio under the action of the form invariance symmetry group. Within this case and due to its  relevance for the inflationary paradigm, we find the general solution of the dynamical equations for a DBI field driven by an exponential potential and show a broad set of inflationary solutions. The general solution
  can be split into three subsets and all these behave asymptotically as a power law solution at early and at late times. 

\end{abstract}

\pacs{98.80.Cq,98.80.Jk,04.20.Jb}
\keywords{brane inflation, DBI theory, form invariance symmetry}
\maketitle

\section{Introduction}

The inflationary paradigm  \cite{23a}-\cite{23d} has been confirmed as the most successful candidate  for explaning the physics of the early universe \cite{6}. This cornerstone of modern cosmology  solves  many of the puzzles of the hot big bang theory, explaning why the universe appears so close to spatial flatness, why it is so homogeneous and isotropic at large scales, and why it is not overwhelmed by magnetic monopoles  \cite{2}. The simplest theoretical framework  to implement inflation is  given by a scalar field  which undergoes a slow roll period responsible for exponential expansion  of the universe at early times. Furthermore, quantum fluctuations at early times provide the seeds for the formation of large-scale structure with a spectrum of adiabiatic, and nearly scale-invariant Gaussian density perturbations. This prediction was confirmed using observational data, mainly those coming from  the cosmic microwave background radiation \cite{observa1a}-\cite{observa1c}, the 2dF galaxy redshift survey \cite{observa2}, and  the SDSS luminous red galaxies  survey \cite{observa3}. 

Recent progresses made within string theory in relation with  the current view of cosmology have pointed out other possible types of physical mechanisms which lead to an inflationary era at early times. For example,  brane inflation tell us  that inflation is entirely due to the motion of D3-brane down a throat in  a warped Calabi-Yau compactification  \cite{pol}; where the central scalar field kinetic energy function is non-canonical and its form is determined by T-duality, yielding  the well known Dirac-Born-Infeld action  \cite{Leigh}. When the spatially homogenous configurations are taken into account the action corresponding to the inflaton field reduces to a DBI field theory which was recently considered in the literature by many authors  \cite{dbi1a}- \cite{dbi1h}. An interesting aspect of this model is a direct consequence of the nonlinear corrections in the kinetic energy. These contributions naturally generate significant levels of primordial non-gaussianity  in the spectrum of curvature perturbations, moreover there is a running of the non-gaussianity reported in  \cite{chen}. In what respect the inflation within the DBI scenario it is important to remark that slow roll can be achieved through a low sound speed instead of dynamical friction due to expansion. Nowadays, the considerable interest devoted to the inflation mechanism with DBI fields has allowed to gain a better understanding of this attractive strings inspired model  \cite{dbi2a}-\cite{dbi2j},  \cite{dbi3a}-\cite{dbi3d}. In this respect, the exact solution  exhibiting power-law inflation with a single DBI field for several  potentials and wraped functions was found in  \cite{dbiL}.


Among the several appealing  scenarios  showing alternative ways to achieve an inflationary stage during the evolution of the Universe,
we are interested in  DBI model. In order to contribute to a deeper understanding of this setup we are using the group of symmetry transformations whereby the Einstein  equations for the spatially flat FRW space time are form invariant  \cite{ai2a}. The group of symmetry is represented by a set of fundamental transformations preserving the form of Friedmann and conservation equations,  mixing geometrical quantities as the expansion rate with quantities of the fluid  as the pressure and the energy density. The aforesaid ``internal'' symmetry leads to a duality between contracting and expanding scenarios, including a phantom stage  \cite{prl}- \cite{nq}. In this context, the cooperative effects  of adding energy density into the Friedmann equation increases the Hubble parameter and the gravitational friction slowing down the rolling of the scalar fields, and in consequence enhancing inflation. Even if a source is not 
energetic enough to provide inflation when the fields are rolling down in isolation, the cumulative effect of a regime of much bigger energy density gives rise to enhanced inflation. This effect was shown to work for tachyon fields  \cite{phta},  interacting fluids  \cite{dualdie,zim}, Chaplygin gases  \cite{dualitycha}, DBI cosmologies  \cite{dbiL},  fermions  \cite{fermion}, quintom scenarios  \cite{nq} brane-world models  \cite{ai2a}- \cite{prl},  DGP cosmologies  \cite{ai3c}- \cite{mariam}, in Bianchi type-V cosmologies  \cite{ai2b} and $n$-dimensional FRW cosmologies  \cite{nd}. 

Other examples of multi-field inflationary models have been explored in the literature  \cite{original}. One can see a resemblance between the enhanced inflation mechanism discussed here and the assisted inflation mechanism, where the cooperative effects  of adding several self-interacting scalar fields driven by an exponential potential make inflation more likely. However, we are going to focus on  enhanced inflation because it works with any fluid and does not require the use of self-interacting scalar fields  or a specific potential. 

In what follows, and as we have anticipated above, we will be concerned with enhanced inflation generated by the subset of linear symmetry transformations. Hence, we will be able to link two cosmological models having different expansion rates, for instance non-accelerated with accelerated ones and vice versa. As will be explained in detail below, in the DBI scenario the key parameter of the linear transformation  of the form invariance symmetry group is given by the Lorentz factor $\gamma$. Below, we will see that in the constant $\ga$ case the Friedmann and conservation equations are fully integrated when the model is driven by an exponential potential.

The paper is organized as follows. In Sec. II we give a review on basic facts of the DBI model. In Subsec. A, following the prescription developed by one of the authors in   \cite{ai2a}, we  investigate the existence of the  form invariance symmetry group within the DBI framework when the group is parametrized by the Lorentz factor $\ga$ and present the transformation rules for physical quantities, such as Hubble expansion rate, the scale factor, the DBI field, the corresponding potential, the barotropic index, the warp factor, the flow parameters of the slow roll approximation, the power spectra for scalar and tensor primordial quantum fluctuations as well as their ratio. We also give  the conditions so that the DBI field drives  an accelerated expansion scenario. In Sec. III we find the general solution for a DBI field driven by an exponential potential with constant Lorentz factor. We present a detailed study of the general solution and show explicitly how enhanced inflation proceeds and how it is induced by the form invariance symmetry group. Finally, in Sec. IV we summarize and discuss our findings. 

\section{DBI enhanced inflation}
Our scenario is that of a four-dimensional spatially flat FRW spacetime 
\be
\n{00}
3H^2=\ro,
\ee
\be
\n{con}
\dot\ro+3H(p+\ro)=0,
\ee
filled with a non-canonical DBI field $\phi$. Here, the dot means differentiation with respect to the cosmological time $t$ and $H=\dot a/a$ is the Hubble expansion rate. Using the customary perfect fluid interpretation $p=(\Ga-1)\ro$, where $\Ga$ is the barotropic index of the DBI field, we have
\be
\n{rp}
\rho=\frac{\ga^2}{\ga+1}\,{\dot\phi}^2+V(\phi)\label{rhodef2},\quad
p=\frac{\ga}{\ga+1}\,{\dot\phi}^2-V(\phi)\label{pdef2},
\ee
\be
\n{g}
\gamma=\frac{1}{\sqrt{1- f(\phi)\dot\phi^2}}, \qquad  \ga>1,
\ee
and 
\be
\n{G}
\Ga=\frac{\ga\dot\phi^2}{\ro},
\ee
where, $f(\phi)>0$ is the warp factor, $V=V(\phi)>0$ is the potential and $\sqrt{f(\phi)}\dot\phi$ may be interpreted as the proper velocity of the brane. Then, for a positive potential, we have that $\Ga$  satisfies the following inequalities
\be
\n{range}
0\le\Ga\le\frac{\ga+1}{\ga}.
\ee

For comparison it is worth mentioning that when the source of the Einstein equations corresponds to a scalar field
with a canonical kinetic term, say a quintessence field $\vphi$, there are different forms to obtain inflationary scenarios. In many of these scenarios the effective potential of a quintessence field is responsible for an epoch of accelerated inflationary expansion. 

Still within the quintessence context, inflationary scenarios  can be obtained also from a more general point of view by using the group of transformations which preserve the form of the Friedmann and conservation equations  \cite{ai2a}. In this case the cooperative effects of adding energy density into the Friedmann equation, instead of quintessence fields, lead different enhanced inflationary scenarios, including phantom one  \cite{prl}. In this interpretation of the inflationary paradigm, the configurations of one and several quintessence fields are related by a simple symmetry transformation of the corresponding Einstein-Klein-Gordon equations. In this sense they can be considered as equivalent cosmological models  \cite{ai2a}.

Now, DBI theory is a framework where the inflation is naturally related with the cooperative effects of adding energy density into the Friedmann equation instead of the cumulative effects of  several DBI  fields. As opposed to the  quintessence field, now the factor  $\ga$ normally increases or magnifies the energy density of the scalar field. This motivation follows from  the inequality $\ro(\ga)\ge\ro_\vphi$, where $\ro_\vphi=\ro(\ga=1)$ is the quintessence field, and the physical assumption that the warp factor  $f(\phi)$ is positive definite. Therefore, the energy density for the DBI field has an increased magnitude as compared with the energy density of the quintessence field. 

\subsection{Form invariance symmetry in DBI theory}

A form invariance transformation is a prescription relating the quantities ($H, \rho, p$) of a FRW cosmology fulfilling Eqs. (\ref{00}-\ref{con}) with the quantities ($\bar{H}, \bar{\rho}, \bar{p}$) of another FRW cosmology described by the Eqs. $3\bar H^2=\bar\rho$ and $\dot{\bar\ro}+3\bar H(\bar p+\bar\ro )=0$. Although the last system of equations have the same form than the original ones (\ref{00}-\ref{con}), both systems represent two different FRW cosmologies because their sources are different. Now we assume that these two cosmologies can be associated with each other by means of the transformation $\bar\ro=\ga^2\ro$. Then the form of the FRW cosmology is then preserved if we complete the transformation as follows
\ben
\n{tr}
&&\bar\ro=\ga^2\ro,\\
\nonumber\\
\n{tH}
&&\bar H=\ga H,\\
\nonumber\\
\n{tG}
&&\bar\Ga=\frac{\Ga}{\ga}-\frac{2\dot{\ga}}{3H\ga^2}.
\een
The latter equation allow us to construct inflationary scenarios ($0<\bar{\Ga}<2/3$) from a non-inflationary one ($2/3<\Ga$) by taking a DBI field that verifies Eq. (\ref{tG}),  we then get
\be
\n{cac}
1-\frac{2\ro}{3\dot{\phi}^2}<\frac{2\dot{\ga} H}{\ga^2\dot{\phi}^2}<1.
\ee

For the simplest  case, $\ga=\mbox{const.}$ with $f\dot\phi^2=\mbox{const.}$, the inequalities (\ref{cac}) are substituted by 
\be
\n{cacs}
\frac{V}{\dot\phi^2}>\frac{3}{2}-\frac{\ga^2}{\ga+1}.
\ee
Now, integrating Eq. (\ref{tH}) we get the relevant transformation rule for the scale factor 
\be
\n{ta}
\bar a=a^\ga.
\ee
The transformation laws for the field, the potential and the barotropic index and are obtained from Eqs. (\ref{rp}), (\ref{G}) and (\ref{tr}) 
\ben
\n{tc}
&&\dot{\bar\phi}^2=\ga\dot\phi^2,\\
\n{tv}
&&\bar V=\ga^2 V+\frac{\ga^3(\ga-1)}{\ga+1}\dot\phi^2,\\
\n{tg}
&&\bar\Ga=\frac{\Ga}{\ga},
\een
where we have assumed that the parameter $\ga$ of the form invariance transformations (\ref{tr}) is an invariant quantity $\bar\ga=\ga$. Hence, $f\dot\phi^2$ is an invariant quantity as well and the warp factor transforms in the same way that the barotropic index (\ref{tG},\ref{tg}) does
\be
\n{tf}
\bar f=\frac{f}{\ga}.
\ee
Finally, taking into account that necessarily $\ga>1$, the DBI  field accelerates the expansion of the universe enhancing inflation when $\ga>3\Ga/2$. 

\subsection{Slow roll approximation} \

A central ingredient of the inflation perturbation is the slow roll approximation. Even when this approximation has not been required by our study of the enhanced inflation in the DBI theory because we have used an internal symmetry of the Friedmann and field equations, it is important to know the transformation rule of the flow parameters under the form invariance transformations to apply it on the power spectra and obtain an overview of the limitations imposed by this approximation. This will set the foundations for the prediction of observational features, and will allow for a critical appraisal of the DBI inflationary scenario.

Following  \cite{dbi3a},  \cite{dbi3c},  \cite{jerarquia}, it can be shown that  the three nonzero fundamental flow parameters, for a constant  Lorentz factor, are given by  
\ben
\n{p1}
&&\epsilon= \frac{2}{\gamma}\left(\frac{H'(\phi)}{H(\phi)}\right)^{2},\\
\n{p2}
&&\eta=\frac{2}{\gamma}\,\frac{H''}{H} ,\\
\n{p3}
&&\zeta=\frac{4}{\gamma^2}\,\frac{H'H'''}{H^2} ,
\een
where the prime  stands for derivates with respect to the fields $\phi$. The slow roll approximation  is valid when $|\epsilon|<<1$, $|\eta|<<1$, and $|\zeta|<<1$ and these requirements impose constraints on the form of the potential and the value of the initial conditions. In the constant $\ga$ case, using the form invariance transformations (\ref{tH}) and (\ref{tc}), one shows that the flow parameter change as 
\ben
\n{tp1}
&&\bar\epsilon= \frac{\epsilon}{\gamma},\\
\n{tp2}
&&\bar\eta=\frac{\eta}{\gamma},\\
\n{tp3}
&&\bar\zeta=\frac{\zeta}{\gamma^2}.
\een

Therefore, we conclude that the slow roll approximation is enhanced when  the energy density is increased in the Friedmann equation.

\subsection{Power spectra}

Density fluctuations generated during the inflationary era are the key ingredient for the formation of large scale structures, and  leave different observational
imprints which may provide a way to constrain inflationary parameters and discriminate among different models. Consequently, we will  be interested in the effect of  the form invariance symmetry on the properties of the spectrum and spectral index for the perturbations that would be created during an inflationary period.

The power spectrum sfor scalar  primordial quantum fluctuation was first derived in  \cite{PertuP} and reads
\be
\n{sp}
{\cal P}_{s}(\tilde k)|_{aH=\tilde k} =\frac{1}{8\pi^{2}}\frac{H^{2}}{\epsilon c_{s}},
\ee
where $c^{2}_{s}=\partial p/\partial \rho=1/\ga^2$ is the speed of sound and the quantity (\ref{sp}) is evaluated  at the time of horizon crossing.
Using (\ref{tH}) and (\ref{tp1}), we find that the power spectrum (\ref{sp}) changes as follows
\be
\n{tsp}
\bar{{\cal P}}_{s}(\tilde k) =\gamma^{3}{\cal P}_{s}(\tilde k).
\ee
The latter result shows that a form invariance transformation, within the DBI model, enhances the power spectrum of the primordial perturbation by
 a factor $\gamma^{3}$. Thus, a DBI cosmological model  with $\gamma>1$ will have naturally  amplified primordial quantum fluctuations. On the other side, the power spectra for the tensor  perturbations are given by 
\be
\n{tp}
{\cal P}_{T}(\tilde k)|_{aH=\tilde k} =\frac{2}{\pi^{2}}H^{2}\propto \rho.
\ee
The tensor power spectrum changes as $\bar{{\cal P}}_{T}=\gamma^{2}{\cal P}_{T}$ under a form invariance symmetry. Then the ratio of power in tensor modes versus scalar modes  \cite{PertuP} 
\be              
\n{rts}
r=\frac{{\cal P}_{T}}{{\cal P}_{s}}, 
\ee
transforms as 
\be
\n{trts}
\bar{r}=\frac{r}{\gamma}, 
\ee
and we find that $\bar{r}<r$  for  $\gamma>1$. Enhanced inflation thus lowers $r$ and makes it easier to meet the $2\sigma$ observational bound
from WMAP7 \cite{wmap7}, which is $r<0.24$ for a power-law power spectrum (actually, this bound gets almost two times larger if
the power-law assumption is relaxed to admit a running index). Future experiments devoted to polarization like CMBPol \cite{CMBPol} 
or Quijote \cite{quij} may offer an improved estimation of this bound.

Another important issue in the DBI model  is related with the non-gaussianities in the perturbation spectra due to its  non-canonical kinetic term. A rough estimator of the non-gaussianity level depends essentially on the Lorentz factor  \cite{dbi1b,dbi1e,jerarquia} 
\be
\n{fNL}
f_{NL}=\frac{35}{108}\Big(\gamma^{2}-1\Big) 
\ee
Then, this measure remains invariant  under the action of the form invariance symmetry. As given by WMAP7  \cite{wmap7}, the current  $2\sigma$ observational bound for non-gaussianity is  $-10 < f_{NL} < 74$,  thus implying $\gamma<15$. Of course, tighter constraints will very likely
be obtained from future missions like those cited above.

\section{The exponential potential}

We start this section by introducing an exponential potential to show explicitly, in this special case, that the  form invariance symmetry leads  to enhanced inflation through the factor $\ga$. For a DBI scenario with $\ga=\mbox{const.}$, after combining Eqs. (\ref{00}-\ref{rp}), we obtain the Friedmann and field equations  
\be
3H^{2}=\frac{\ga^2}{\ga+1}\,{\dot\phi}^2+V, \label{ec.f}
\ee
\be
\ddot\phi+\frac{3(\ga+1)}{2\ga}\,H\dot\phi+(\ga+1)\frac{V'}{2\ga^2}=0, \label{feq}
\ee
with $V'=dV/d\phi$, and the useful equations
\be
-2\dot H=\ga{\dot\phi}^2, \label{mastereq}
\ee
\be
3H^2+\frac{2\ga}{\ga+1}\dot{H}=V.\label{consiste}
\ee

A simple realization of  the condition (\ref{cacs}) is to assume that $V(\phi)\propto \dot\phi^2$ (with an appropriate proportionality
factor to be determined later). Hence, $V'\dot\phi\propto \dot\phi\ddot\phi$, $V'\propto \ddot\phi$ and from Eq. (\ref{ec.f}) we get $H\propto V^{1/2}$. Inserting all these results in the equation field (\ref{feq}) we obtain $V'\propto V$, meaning that the DBI field is driven by an exponential potential. Therefore, we will write  the potential  as
\be
V=V_{0}\,{\mbox e}^{-B\phi}\label{Vdef},
\ee
where $V_0$ and $B$ are constants with $V_0>0$.

The exact solution of the Friedmann and field equations (\ref{ec.f}-\ref{feq}) for the exponential potential (\ref{Vdef}) (so called Liouville potential) includes power law solutions, a particular explicit solution depending on the cosmological time and the general solution which implicitly depends  on a function of the cosmological time. All these solutions have the common feature of a  power law asymptotic behavior. 

To proceed we introduce an auxiliary  function of the cosmological time $L(t)$ and rewrite $\dot\phi$ in terms of this new variable as
\be
\dot\phi=\frac{B}{\ga}\,H+L.\label{fint}
\ee
This becomes a first integral of the DBI  field equation (\ref{feq}) whenever the function $L$ satisfies the equation 
\be
\dot L + \frac{3H(\ga+1)}{2\ga}L=0.
\ee
By inserting the solution of the last equation $L=b~a^{-3(\ga+1)/2\ga}$ into  Eq. (\ref{fint}), we get the first integral of the field equation (\ref{feq}) 
\bn{fi}
\dot\phi=\frac{B}{\ga}\,H+b~a^{-\frac{3(\ga+1)}{2\ga}},
\ee
where $b$ is an integration constant. Thus, the Liouville potential (\ref{Vdef}) helps us to find the first integral  of the field equation by turning the latter  into a separable non-linear differential equation. The separation  yields two parts: one depends exclusively on geometrical quantities,  whereas the  other depends on quantities that characterize the field.

Replacing the first integral (\ref{fi}) into the Friedmann equation (\ref{ec.f}), we get a quadratic equation in the expansion rate $H$ 
\be
\n{frix}
\left[3(\ga+1)-B^2\right]H^2-\frac{2\ga bBH}{a^{3(\ga+1)/2\ga}}-\frac{\ga^2 b^2}{a^{3(\ga+1)/\ga}}-V=0,
\ee
whose discriminant $D$ must be definite positive to obtain real valued scale factors, so we find the following inequality:
\be
\n{dpos}
[3(\ga+1)-B^2]V+\frac{3\ga^2b^2}{a^{3(\ga+1)/\ga}}>0.
\ee
Note that the coefficient of the potential must be positive so that the inequality (\ref{dpos}) holds in the limit of large scale factor, so one concludes that the restriction
\be
\n{range'}
B^2< 3(\ga+1),
\ee
must hold in order to have real valued solutions with non null potential.

\subsection{Power law solutions}

For the particular $b=0$ case the first integral (\ref{fi}) becomes $\dot\phi=BH/\ga$, so by integration we get $\phi=(B/\ga)\ln{a}$, where the integration constant has been set to zero. On the other hand by combining Eqs. (\ref{ec.f}), (\ref{feq}) and (\ref{Vdef}) the field equation (\ref{feq}) reduces to
\bn{kg1}
\ddot\phi+\frac{B}{2}\dot\phi^2=0.  
\ee
After integrating twice the Eq. (\ref{kg1}), we get the DBI field, the scale factor from $a=\exp(\ga\phi/B)$ and $V_0$ from the Friedmann equation (\ref{ec.f}), they read  
\be
\n{todos}
a=t^{2\ga/B^2}, \qquad \phi=\frac{2}{B}\ln{t},  
\ee
\be
\n{v0}
V_0=\frac{4\ga^2}{B^2}\left[\frac{3}{B^2}-\frac{1}{\ga+1}\right]. 
\ee
The DBI model enhances inflation for any value of $B$ such that $\ga>B/2$. In a way, for arbitrary $\ga$ values the enhanced framework generalizes the assisted inflation. In fact, by applying the form invariance transformation (\ref{tr}), (\ref{tH}) and (\ref{ta}) on the power law solution $a=t^{2/B^2}$, we obtain $\bar a=a^\ga=t^{2\ga/B^2}$. It means that the transformed cosmological DBI model is filled with an energy density $\bar\ro$ larger than the energy density $\ro$ of the original model, because $\bar\ro=\ga^2\ro$ with $\ga>1$. This shows explicitly  that the cooperative effects of adding energy density into the Friedmann equation enhances inflation. Similarly the barotropic index $\Ga$ of the DBI field transforms as $\bar\Ga=\Ga/\ga$ under a form invariance transformation enhancing inflation for $\ga>3\Ga/2$ . 

The particular power law solution $a=t^{2\ga/3(\ga+1)}$ is excluded for the exponential potential because in this case $B^2=3(\ga+1)$ and $V_0$ vanishes in Eq. (\ref{v0}).  

\subsection{Particular explicit solution}

In order to obtain some particular explicit solutions of Eqs. (\ref{ec.f}), (\ref{feq}) and (\ref{Vdef}), we combine Eqs. (\ref{mastereq}), (\ref{fi}) which  lead us  to a nonlinear second order differential equation for the scale factor $a$:
\be
-2\dot{H}=\ga \left[\frac{B^2H^2}{\ga^2}+\frac{2bBH}{\ga}\,a^{-\frac{3(\ga+1)}{2\ga}}+b^2a^{-\frac{3(\ga+1)}{\ga}}\right]. \label{aeq1}
 \ee
By making the change of variables 
\begin{equation}
s=a^{B^2/2\ga }, \qquad \tau=bB\,t, \qquad m=-\frac{3(\ga+1)}{B^{2}},\label{cambioent}
\end{equation}
it follows from Eq. (\ref{range'}) that $m\le -1$, hence the Eq.(\ref{aeq1}) becomes a second order nonlinear differential equation for $s(\tau)$:
\begin{equation}
\ddot s + s^{m}\dot s+\frac{1}{4}\,s^{2m + 1}=0,\label{NL}
\end{equation}
where the dot stands for differentiation with respect to the argument of the function.  Once we have found the solution of Eq. (\ref{NL}), we will obtain the ``scale factor $a$'' in terms of the cosmic time $t$ by using Eqs. (\ref{cambioent}). After that we must constraint the ``scale factor $a$'' in order for it   to be a comply with the  Friedmann equation (\ref{ec.f}) also. This will allow us to bind the two integration constants, provided that the two integrations of Eq. (\ref{NL}) are related, so that finally we have the true scale factor with only one relevant integration constant. 

By making a change of variable $s(\tau)\to v(\tau)$ 
\bn{ex}
\dot s=\al s^{m+1}+s\frac{\dot v}{v},
\ee
where $\al$ is a constant to be determined later. Under this new variable  Eq. (\ref{NL}) can be recast as 
\bn{et}
\left[(m+1)\al^2+\al+\frac{1}{4}\right]s^{2m+1}+\left[(m+2)\al+1\right]\frac{\dot v}{v}+s\frac{\ddot v}{v}=0.
\ee
For the two values, $m=0$ with $\al=-1/2$ and $m=-4$ with $\al=1/2$, the last equation reduces to 
\bn{..}
\ddot v=0.
\ee
We discard  the $m=0$ value because it entails $B^2\to\infty$ and a vanishing exponential potential (\ref{Vdef}). However, for  $m=-4$, we integrate Eq. (\ref{ex}) and obtain 
\begin{equation}
s^{-4}=\frac{v^{-4}}{2\int{v^{-4}d\tau}}.\label{sustitu}
\end{equation}
By solving the latter with $v$ given by Eq. (\ref{..}), i.e, $v=b_{1}+b_{2}\tau$, where $b_1$ and $b_2$ are constants, and substituting into the first Eq. (\ref{cambioent}), we  obtain the scale factor 
\begin{equation}
a=\left[\frac{2\left(k\tau^{4}- \tau\right)}{3}\right]^{2\ga/3(\ga+1)}.\label{a(v)}
\end{equation}
Without loss of generality, the quantity $(k\tau^{4}-\tau )$ must be restricted to remain positive definite, so we take $b_1=0$ and the constant $k$ becomes a redefinition of the old one. 
 
The DBI field $\phi(\tau)$ is obtained by integrating the first integral (\ref{fi}) of the field equation (\ref{feq}) 
\begin{equation} 
\n{fie}
\phi=\frac{1}{B}\ln{\left[\frac{\left|\phi_{0}\right|}{3}\,\tau^{2}(k\tau^{4}-\tau)\right]},
\end{equation}
where $\phi_{0}$ is an integration constant. Finally, from Eqs. (\ref{Vdef}) and (\ref{fie}), we find the exponential potential as a function of the cosmological time
\bn{pot}
V=\frac{3V_0\tau^2}{\left|\phi_{0}\right|(k\tau^{4}-\tau)}.
\ee

To constrain the integration constants we use the scale factor (\ref{a(v)}) to calculate $H$, $\dot H$ and insert them along with the potential (\ref{pot}) into the Friedmann equation (\ref{consiste}). The dependence on the variable $v$ is missed and we find a relation between $b$, $\phi_0$, $k$ and $V_0$:
\begin{equation}
k=\frac{V_{0}(\ga+1)}{4|\phi_{0}|b^2\ga^{2}}.\label{relaconstantes}
\end{equation}
This gives a positive definite constant $k$. Hence, the scale factor (\ref{a(v)}) and the DBI field (\ref{fie}), whose integration constants verify the relation (\ref{relaconstantes}), are exact solutions of the Friedmann and field equations (\ref{ec.f})-(\ref{feq}) for the exponential potential (\ref{pot}).

\subsection{Implicit general solution}

Motived by the use of the exponential potential  in the framework DBI inflation framework, we will find the general implicit solutions of Eqs. (\ref{ec.f}), (\ref{feq}) and (\ref{Vdef}). To this end, we will solve  Eq. (\ref{NL}) by using the  form invariance group generated by a non-local transformation  of variables acting on the nonlinear differential equation 
\begin{equation}
\ddot{s} +F\dot{s}+ \beta~F\int{Fds}+\nu F=0\label{IF1}, 
\end{equation}
where $s=s(\tau)$, $F=F(s)$ is any integrable real function, the dot means differentiation with respect to $\tau$ whereas $\beta$ and $\nu$ are constant parameters. In our DBI model Eq. (\ref{NL}) is obtained from Eq. (\ref{IF1}) by selecting $F=s^n$. Below, we will use a non-local transformation of variables to linearize Eq. (\ref{IF1}).  

We introduce a new pair of variables $(h,\eta)$ by means of the variable transformations  
\begin{eqnarray}
F(s)ds=G(h)dh,\n{n1}
\qquad 
F(s)d\tau=G(h)d\eta, 
\end{eqnarray}
where $G(h)$ is an integrable real function. The integration of the first transformation in Eq. (\ref{n1}) defines $h=h(s)$. However, the second transformation in Eq. (\ref{n1}) cannot be integrated until we know the functions $F(s)$ and $G(\eta)$. For this reason it defines a non-local variable transformation. Although, we do not know it explicitly, it transforms the Eq. (\ref{IF1}) into a nonlinear ordinary differential equation having the same form,
\begin{equation}
h''+G h' + \beta G\int{G dh}+\nu G=0\label{IF2}, 
\end{equation}
where $'$ denotes differentiation with respect to $\eta$. In other words Eqs. (\ref{IF1}) and (\ref{IF2}) are related between them by the formal changes $F\leftrightarrow G$, $s\leftrightarrow h$ and $\tau\leftrightarrow\eta$. Then, the non-local transformation (\ref{n1}) preserves the form of these equations mapping solutions of  Eq. (\ref{IF1}) into solutions of Eq. (\ref{IF2}) for any pair of functions $F(s)$ and $G(\eta)$, linking solutions of two different physical configurations. Taking into account the nature of the non-local change of variables (\ref{n1}), it is not always possible to find explicit solutions of the Eq. (\ref{IF1}). However, the procedure based in the use of the form invariance relates the class of nonlinear differential equations (\ref{IF1}), generated by the function $F(s)$, with the linear damped harmonic oscillator equation, by selecting $G(h)=const.$. In our case, we apply this procedure by choosing the specific parametrization $F(s)=s^{m}$ an
 d $G(h)=1$ in  Eqs. (\ref{IF1}) and (\ref{IF2}). These two different configurations are described by the equations
\be
\ddot{s}+s^{m}\dot{s}+ \frac{\beta}{m+1}\,s^{2m+1}+\nu s^m=0,\n{s''}
\ee
\be
h''+h' + \beta h +\nu=0, \n{h}
\ee
and their solutions are related by the non-local variable transformations (\ref{n1}) evaluated on the above specific parametrization  
\be
h=\int{s^{m}ds},\n{c1}
\qquad
\eta=\int{s^{m}d\tau}.
\ee

Summarizing, the non-local change of variables (\ref{c1}) transforms the non-linear differential equation (\ref{s''}), describing a DBI field driven by an exponential potential, into a linear damped  harmonic oscillator equation (\ref{h}) with a constant gravitational field. It means that the non-local change of variables (\ref{c1}) alters the Lie point symmetry of these two physical problems making them essentially the same in the new variables ($h$,$\eta$).

Comparing our equation (\ref{NL}) with Eqs. (\ref{s''}) and (\ref{h}), we have two different cases to analyze,
\begin{itemize}
	\item $m\neq -1$,\,\,\,\,\,\, $\nu=0$,\,\,\,\,\,\, $F(s)=s^{m}$,\,\,\,\,\,\,$\beta={(m+1)}/{4}$
	\item $m~= -1$,\,\,\,\,\,\,$\beta=0$,\,\,\,\,\,\, $F(s)=
		{1}/{s}$,\,\,\,\,\,\,\,\,\,$\nu={1}/{4}$
\end{itemize}	
\no For which the Eq. (\ref{NL}) becomes 
\be
\n{hh}
h''+h'+\frac{m+1}{4}h=0, \qquad  m\neq -1,
\ee
\be
h''+h'+\frac{1}{4}=0.  \qquad  m=-1. \n{h'}
\ee

For $m\neq-1$ and $\nu=0$, we obtain from Eqs. (\ref{Vdef}), (\ref{fi}), (\ref{cambioent}), (\ref{c1}) and (\ref{hh})  the scale factor, the DBI field and the potential

\begin{equation}
a(\eta)=\left[(m+1)\left(c_{1}{\mbox e}^{\lambda^-\eta}+ c_{2}{\mbox e}^{\lambda^{+}\eta}\right)\right]^{\frac{2\ga }{B^2-3(\ga+1)}},\label{aparamet}
\end{equation}

\begin{equation}
\n{pi}
\phi(\eta)=\frac{1}{B}\left[\ln{|\phi_{0}|}+\eta+\frac{B^2}{\ga}\ln a\right],
\end{equation}

\be
\n{pf}
V=\frac{V_0\,{\mbox e}^{-\eta}}{|\phi_0|\, a^{B^2/\ga}},
\ee
where $\lambda^{\pm}=(-1 \pm \sqrt{-m})/2$ are the roots of the characteristic polynomial of the Eq. (\ref{hh}) and $\phi_{0}$ is an integration constant. Now, inserting Eqs. (\ref{aparamet}), (\ref{pi}) and (\ref{pf}) into the Friedmann equation (\ref{consiste}), the integration constants are constrained so they satisfy the following relation:
\begin{equation}
c_{1}c_{2}=\frac{B^4 V_{0}}{12\ga^2 b^2[B^2-3(\ga+1)]|\phi_{0}|}.\label{consistenciaparam1}
\end{equation}
Defining the constant $C^2=-4c_1c_2$ and taking into account Eqs. (\ref{cambioent}), (\ref{aparamet}) and (\ref{consistenciaparam1}), we obtain the final form of the scale factor for $B^2-3(\ga+1)<0$,
\be
\n{af}
a=\left[C\,{\mbox e}^{-\eta/2}\sinh{\left(\frac{\sqrt{3(\ga+1)}}{2B}\,\eta-\sigma\right)}\right]^{\frac{2\ga }{B^2-3(\ga+1)}}.
\ee

At early times, $\eta\to\infty$, the DBI model begins  at a singularity where the scale factor reads
\be
\n{age}
a_e\approx {\mbox e}^{2\la^{+}\ga\Delta\eta/[B^2-3(\ga+1)]}.
\ee
The relation between the implicit time $\eta$ and the cosmological time is given by the second Eq. (\ref{c1}),
\be
\n{tge}
\Delta\tau\approx {\mbox e}^{3\la^{+}(\ga+1)\Delta\eta/[B^2-3(\ga+1)]},
\ee
then, composing the last two equations we obtain the initial behavior of the implicit solutions
\be
\n{aef}
a_e\approx\Delta\tau^{2\ga/3(\ga+1)}.
\ee

At late times the constant $\sigma$ defines the limit value of the implicit time $\eta_\infty =2B\sigma/\sqrt{3(\ga+1)}$ where the scale factor reaches $a=\infty$. To find the behavior of the latter set of solutions for a large scale factor, we expand it near $\eta_\infty$ by making $\eta=\eta_\infty+\Delta\eta$ with $\Delta\eta>0$ and $\Delta\eta/\eta_\infty\ll 1$. In this limit, the scale factor behaves as
\be
\n{ainf}
a_l\approx \Delta\eta^{\frac{2\ga }{B^2-3(\ga+1)}},
\ee
while from the second Eq. (\ref{c1}), the cosmological time becomes
\be
\n{tinf}
\Delta\tau\approx \Delta\eta^{\frac{B^2}{B^2-3(\ga+1)}}.
\ee
Finally, from the last two we get 
\be
\n{ainff}
a_l\approx \Delta\tau^{2\ga/B^2}.
\ee
Eqs. (\ref{aef}) and (\ref{ainff}) show  that the DBI model driven by an exponential potential interpolates between two fluids with barotropic indexes $\Ga_e=(\ga+1)/\ga$ and $\Ga_l=B^2/3\ga$. Actually the barotropic index of the DBI field becomes variable and it can be calculated by inserting the exact scale factor (\ref{af}) into $\Ga=-2\dot H/3H^2$, hence the final result is 
\be
\n{Gamma}
\Ga=\frac{\ga+1}{\ga}\left[\frac{B-\sqrt{3(\ga+1)}\tanh{\frac{\sqrt{3(\ga+1)}}{2B}\,\eta-\sigma}}{\sqrt{3(\ga+1)}-B\tanh{\frac{\sqrt{3(\ga+1)}}{2B}\,\eta-\sigma}}\right]^2.
\ee
 
For the $m= -1$ ($B^2 =3(\ga+1)$), $\beta=0$ and $\nu=1/4$ case we have
\begin{eqnarray}
&&a(\eta)=\exp{\left[\frac{2\ga}{3(\ga+1)}\left(d_1+d_2\,{\mbox e}^{-\eta}-\frac{\eta}{4}\right)\right],\label{aparamet2}}
\\
&&\phi(\eta)=\frac{1}{B}\left[\ln{|\phi_{0}|}+\eta+\frac{B^2}{\ga}\,\ln{a}\right].\label{phiparame2}
\end{eqnarray}
Combining Eqs. (\ref{consiste}), (\ref{pf}), (\ref{aparamet2}) and (\ref{phiparame2}), the integration constants  should fulfill the following constraint:
\be
\n{d2}
d_{2}=\frac{(\ga+1)V_{0}}{4\ga^2 b^2|\phi_{0}|} \label{consparame2}. 
\ee
Hence the final scale factor can be rewritten  as
\be
\n{a-1}
a(\eta)=a_0\exp{\left[-\frac{\ga\eta}{6(\ga+1)}+\frac{V_0\,{\mbox e}^\eta}{6\ga b^2|\phi_{0}|}\right]},
\ee
where $a_0$ is a redefinition of the constant $d_1$.

Summarizing, the general solution of the a DBI field  driven by an exponential potential and constant factor $\ga$ is composed by scale factors (\ref{todos}), (\ref{a(v)}), (\ref{af}), (\ref{a-1}) and the corresponding DBI fields (\ref{todos}), (\ref{fie}), (\ref{pi}), (\ref{phiparame2}). Analyzing the asymptotic limits of the scale factor  (\ref{af}) we see that $a_e\to t^{2\ga/3(\ga+1)}$ at early times and $a_l\to t^{{2\ga/B^2}}$ at late times. It transits from $a_e\approx t^{1/3}$ and  $a_l\approx t^{2/B^2}$ for $\ga\to 1$ (quintessence field) to $a_e\approx t^{2/3}$ and $a_l\approx t^{{2\ga/B^2}}$ for $\ga\gg 1$ (DBI  field), showing that the exponents increase from $1/3$ and (${\ga/B^2}$) to $2/3$ and (${2\ga/B^2}$) at early times and (late times) respectively. Therefore, a DBI field driven by an exponential potential does not inflate near the initial singularity. However, away from the singularity, when the time proceeds, the inflation becomes more probable and it is enhanced by the factor $\ga$ within the DBI framework.

\section{Conclusion}

By using the very appealing form invariance symmetry group contained in the Friedmann and conservation equations, we have given a full description of the enhanced inflation within the DBI cosmological model for spatially flat FRW spacetimes and we have shown that the cooperative effects of increasing the energy density in the Friedmann equation makes  inflation more likely. We have increased the energy density of the DBI field by means of the form invariance transformation $\ro=\ga^2\ro$, through the ``real Lorentz factor'' $\gamma>1$. It substitutes the configuration of $n$ self-interacting DBI fields,  contrasting with the assisted inflation originally derived for the  quintessence field. In the peculiar case that the  Lorentz factor $\gamma$ is a natural number, it plays the same role that the number of $n$ self-interacting DBI fields. Hence, the DBI field accelerates the expansion of the universe, leading to an inflationary era, when its  barotropic index $\Gamma$ and the Lorentz factor $\ga$
  satisfy the inequality $\gamma>3\Gamma/2$. 

In this context we have also investigated some interesting examples by determining the transformation rules for the scalar and tensor power spectra, as well as their ratio, and the slow roll parameters under the action of the form invariance transformations. We have found that  the primordial quantum fluctuations are naturally amplified by $\ga$ and the slow roll approximation is enhanced when  the energy density is increased in the Friedmann equation.

We have obtained the general solution of the Friedmann and DBI field equations for a constant Lorentz factor when the DBI field is driven by an exponential potential. The general solutions has been split into three subsets, power law solutions, an explicit particular solutions and the implicit general solution. We have shown that they all share the property that at early and at late times they behave asymptotically as power law solutions. The first subset of solutions exhibit enhanced inflation under a form invariance transformation. On the other hand, the asymptotic behavior of the remaining two subsets of solutions  at late times indicates that  inflation becomes more probable and it is enhanced by the $\ga$ factor.\\
 
\acknowledgments

LPC thanks the University of the Basque Country, the Basque Foundation for Science (Ikerbasque), the University of Buenos Aires under Project No. X044 and the Consejo Nacional de Investigaciones Cient\'{\i}ficas y T\' ecnicas (CONICET) under Project PIP 114-200801-00328 for the partial support of this work during its different stages. MGR is supported by CONICET. RL thanks the Spanish Ministry of Science and Innovation and the University of the Basque Country for support through research projects FIS2010-15492 and GIU06/37 respectively. 



\begin{thebibliography}{99}
\bibitem{23a}
A. A. Starobinsky, JETP Lett., {\bf 30}, 682 (1979).
\bibitem{23b}
 A. H. Guth, Phys. Rev. D {\bf 23}, 347 (1981).
\bibitem{23c}
A. D. Linde, Phys. Lett. B {\bf 108}, 389 (1982).
\bibitem{23d}
A. Albrecht and P. J. Steinhardt, Phys. Rev. Lett. {\bf 48},
1220 (1982).
\bibitem{6}
W. H. Kinney, E. W. Kolb, A. Melchiorri,  and A. Riotto,
Phys. Rev. D {\bf78}, 087302 (2008).
\bibitem{2}
A. R. Liddle and D. H. Lyth, \emph{Cosmological Inflation
and Large-scale Structure}, (Cambridge : Cambridge UP,
2000), p. 400.
\bibitem{observa1a}
P. de Bernardis et al., Astrophys. J. {\bf564}, 559 (2002).
\bibitem{observa1b}
C. Pryke et al., Astrophys. J. {\bf568}, 46 (2002).
\bibitem{observa1c}
E. Komatsu et al, Astrophys. J. Suppl.Ser., {\bf180}, 330
(2009).
\bibitem{observa2}
S. Cole et al., Mon. Not. Roy. Astron.
Soc. {\bf362}, 505 (2005). 
\bibitem{observa3}
M. Tegmark et al., Phys. Rev. D {\bf74}, 123507
(2006). 

\bibitem{pol}
S. B. Giddings, S. Kachru and  J. Polchinski, Phys. Rev. D  {\bf66}, 106006 (2002).
\bibitem{Leigh}
R. G. Leigh, Mod. Phys. Lett. A {\bf4}, 2767 (1989).
\bibitem{dbi1a}
E. Silverstein and D. Tong, Phys. Rev. D {\bf 70}, 103505 (2004).
\bibitem{dbi1b}
M. Alishahiha, E. Silverstein and D. Tong, Phys. Rev. D {\bf70}, 123505 (2004).
\bibitem{dbi1c} 
 J. E. Lidsey and D. Seery, Phys. Rev. D {\bf75}, 043505 (2007).
\bibitem{dbi1d}
D. Baumann and L. McAllister, Phys. Rev. D {\bf75}, 123508 (2007).
\bibitem{dbi1e}
 X. Chen, M. x. Huang, S. Kachru and G. Shiu, JCAP, {\bf0701}, 002 (2007).
\bibitem{dbi1f}
  R. Bean, S. E. Shandera, S. H. Henry Tye and J. Xu, JCAP {\bf 0705}, 004 (2007).
\bibitem{dbi1g}
M. Spalinski, JCAP {\bf0704}, 018 (2007).
\bibitem{dbi1h}
M. Spalinski, JCAP {\bf 0705}, 017 (2007).
\bibitem{chen}
X. Chen, Phys. Rev. D {\bf72}, 123518 (2005).


\bibitem{dbi2a}
S. Mizuno, F. Arroja and K. Koyama, Phys. Rev. D  {\bf80}, 083517 ( 2009).
\bibitem{dbi2b}
W. H. Kinney and  K. Tzirakis, Phys. Rev. D  {\bf77},103517 (2008).
\bibitem{dbi2c}
S. Mizuno, F. Arroja, K. Koyama and  T. Tanaka, Phys. Rev. D  {\bf80}, 023530 (2009).
\bibitem{dbi2d}
D.Langlois, S. Renaux-Petel and D. A. Steer, JCAP {\bf0904}, 021  (2009).
\bibitem{dbi2e}
D. Langlois, S. Renaux-Petel, D. A. Steer and T. Tanaka, Phys. Rev. D  {\bf78}, 063523 (2008).
\bibitem{dbi2f}
K. Bamba, N. Ohta and S. Tsujikawa, Phys. Rev. D  {\bf78}, 043524 (2008).
\bibitem{dbi2g}
A. J. Tolley and  M. Wyman, JCAP {\bf0804}, 028 (2008).
\bibitem{dbi2h}
D.A. Easson, R. Gregory, D. F. Mota, G. Tasinato, and I. Zavala, JCAP {\bf0802}, 010 (2008).
\bibitem{dbi2i}
T. Kobayashi, S. Mukohyama and S. Kinoshita, JCAP {\bf0801}, 028 (2008).
\bibitem{dbi2j}
M. Spalinski, Phys. Lett. B {\bf650}, 313 (2007).
\bibitem{dbi3a}
D. Bessada, W. H. Kinney, K.Tzirakis, JCAP {\bf0909}, 031,2009.
\bibitem{dbi3b}
W. H. Kinney  and K. Tzirakis,  Phys. Rev. D  {\bf77}, 103517 (2008).
\bibitem{dbi3c}
M. Spalinski, JCAP {\bf0804}, 002 (2008).
\bibitem{dbi3d}
M. Spalinski, JCAP {\bf0705}, 017 (2007).
\bibitem{dbiL}
L. P. Chimento and R. Lazkoz, Gen. Rel. Grav. {\bf40}, 2543 (2008).
\bibitem{ai2a}
L. P. Chimento, Phys. Rev. D {\bf65}, 063517 (2002).
\bibitem{prl}
L. P. Chimento and  R. Lazkoz,
Phys. Rev. Lett. {\bf 91}, 211301 (2003).
\bibitem{phta}
J.M. Aguirregabiria, L.P. Chimento,  and R. Lazkoz, 
Phys. Rev. D {\bf 70}, 023509, 2004.
\bibitem{uniphan} 
L.P. Chimento and R. Lazkoz, 
Int. J. Mod. Phys. D {\bf 14}, 587 (2005).

\bibitem{dualdie}
L. P. Chimento and Diego Pav\'on, Phys. Rev. D {\bf 73}, 063511 (2006). 

\bibitem{dualitycha}
L.P. Chimento and R. Lazkoz,
Class. Quant. Grav. {\bf 23}, 3195 (2006). 

\bibitem{zim}
L. P. Chimento and W. Zimdahl,
Int. J. Mod. Phys. D {\bf 17}, 2229 (2008). 


\bibitem{fermion}
L. P. Chimento, F. P. Devecchi, M. I. Forte and G. M. Kremer,
Class. Quant. Grav. {\bf 25}, 085007 (2008). 

\bibitem{nq}
L. P. Chimento, M. I. Forte, R. Lazkoz and M. G. Richarte,
Phys. Rev. D  {\bf79}, 043502 (2009). 

\bibitem{ai3c}
R. Lazkoz, Phys. Rev. D  {\bf70}, 064033 (2004). 

\bibitem{mariam}
M. Bouhmadi-L\'opez, L.P. Chimento, arXiv:1007.4141.

\bibitem{ai2b} 
J. M. Aguirregabiria,  L. P. Chimento, A. S. Jakubi and  R. Lazkoz,
Phys. Rev. D  {\bf67}, 083518  (2003). 

\bibitem{nd}
M. Cataldo and L.P. Chimento,
Int. J. Mod. Phys. D {\bf17}, 1981 (2008). 





\bibitem{original}
A. R. Liddle, A. Mazumdar and F. E. Schunck, Phys. Rev. D {\bf58}, 061301 (1998);
E.J. Copeland, A. Mazumdar A and N.J. Nunes,  Phys. Rev. D {\bf60}, 083506 (1999); 
P. Kanti  and K.A. Olive, Phys. Rev. D {\bf60}, 043502 (1999);  
P. Kanti  and K.A. Olive, Phys. Lett. B {\bf464}, 192 (1999);  
N. Kaloper and A.R. Liddle, Phys. Rev. D {\bf61}, 123513 (2000);  
A. Mazumdar,  S. Panda, A. P\'erez-Lorentzana, Nucl. Phys. B {\bf614}, 101 (2001);
Y.S. Piao, W.b. Lin, X.m. Zhang,  Y.Z. Zhang, Phys. Lett. B {\bf528}, 188 (2002);
Y.S. Piao et al., Phys. Rev. D {\bf66} 121301  (2002); 
R. Brandenberger, P.M. Ho and H.c. Kao, JCAP {\bf0411}, 011 (2004); 
A. Jokinen  and A. Mazumdar, Phys. Lett. B {\bf597}, 222  (2004); 
K. Becker, M. Becker and A. Krause, Nucl. Phys. B {\bf 715}, 349 (2005);
J.D. Barrow and N.J. Nunes, Phys. Rev. D  {\bf 76} 043501 (2007); 
K. L. Panigrahi and  H. Singh, JHEP {\bf0711} 017 (2007);
S. Dimopoulos, S. Kachru, J. McGreevy and  J. Wacker, JCAP {\bf0808} 003 (2008);



\bibitem{PertuP}
J. Garriga and V. F. Mukhanov, Phys. Lett. B{\bf458}, 219 (1999).
\bibitem{F}
L. P. Chimento 
J. Math. Phys. {\bf 38}, 2565-2576 (1997).
\bibitem{wmap7}
E. Komatsu et al. 2010, arXiv:1001.4538.
\bibitem{jerarquia}
H. V. Peiris , D. Baumann, B. Friedman,A. Cooray, Phys. Rev. D {\bf 76} 103517 (2007). 
\bibitem{CMBPol}
D. Baumann et al., AIP Conf.Proc. {\bf 1141} 10 (2009).
\bibitem{quij}
J.~A.~Rubino-Martin et al., arXiv:0810.3141.
\end{thebibliography}
\end{document}